\documentclass[aip,amsmath,amssymb,reprint,]{revtex4-1}
\usepackage{graphicx}
\usepackage{dcolumn}
\usepackage{bm}
\usepackage[utf8]{inputenc}
\usepackage[T1]{fontenc}
\usepackage{mathptmx}

\newcommand{\be}{\begin{equation}}
\newcommand{\ee}{\end{equation}}

\begin{document}
\preprint{AIP/123-QED}

\title[Preprint submitted to Physics of Plasmas]{Seed Source for Plasma Compression in the Long Wavelength Infrared}

\author{D.F. Gordon}
\affiliation{Naval Research Laboratory, Plasma Physics Division, Washington, DC 20375, USA}
\author{P. Grugan}
\altaffiliation{National Research Council Research Associate}
\affiliation{Naval Research Laboratory, Plasma Physics Division, Washington, DC 20375, USA}
\author{R. Kupfer}
\affiliation{Brookhaven National Laboratory, Accelerator Test Facility, Upton, NY}
\author{Y.-H. Chen}
\affiliation{Naval Research Laboratory, Plasma Physics Division, Washington, DC 20375, USA}
\author{A. Ting}
\affiliation{Commonwealth Technology Innovation LLC, Alexandria, VA 22315}
\author{A. Mamonau}
\affiliation{Naval Research Laboratory, Plasma Physics Division, Washington, DC 20375, USA}
\author{L.A. Johnson}
\affiliation{Naval Research Laboratory, Plasma Physics Division, Washington, DC 20375, USA}
\author{M. Babzien}
\affiliation{Brookhaven National Laboratory, Accelerator Test Facility, Upton, NY}

\date{\today}

\begin{abstract}
Two color laser pulses are used to form an air plasma and generate broadband infrared radiation suitable as a seed for backward Raman amplification of CO$_2$ laser pulses.  Broadband radiation in the atmospheric window from 8-14 microns is observed.  The infrared radiation is characterized using a long wavelength grating spectrometer specially designed to accept an ionizing laser filament at its input plane.  The LWIR yield is greatly enhanced by chirping the drive pulse.  Unidirectional pulse propagation simulations suggest that this is due in part to the dependence of the nonlinear refractive index on the pulse duration.
\end{abstract}

\maketitle

\section{Introduction}

Sources of long wavelength infrared (LWIR) radiation with terawatt peak power are currently limited to highly specialized CO$_2$ laser systems \cite{haberberger2010,polyanskiy2017}.  Recently, $\sim 100$ gigawatt single-cycle pulses were obtained via down-conversion of near infrared ultra-short pulses in a relativistic plasma \cite{nie2020}.  This paper is directed toward producing sub-picosecond, terawatt LWIR pulses, by means of backward Raman amplification (BRA) \cite{malkin1999,ping2004,wu2016,turnbull2018}.  In general, BRA is a form of plasma compression, requiring a high energy pump pulse, and a low energy, counter-propagating seed pulse.  This process is especially useful for compressing CO$_2$ laser pulses, because the limited bandwidth of the pump pulse does not affect the final compression ratio.  Analysis and simulations of long wavelength BRA are presented in Ref.~\cite{johnson2017}.

To compress a CO$_2$ laser pulse via BRA requires an ultra-short seed pulse with wavelength $\lambda\gtrsim 10$ $\mu$m.  The energy of the seed pulse is not a critical parameter, but for alignment and synchronization purposes a pulse energy of $\sim 1$ $\mu$J is desirable.  To lowest order, the center frequency of the seed pulse should be $\omega_s = \omega_0 - \omega_p$, where $\omega_0$ is the pump frequency and $\omega_p$ is the plasma frequency.   Changing the plasma density, and therefore $\omega_p$, tunes the frequency of the compressed pulse.  The most convenient seed source would either be tunable with modest bandwidth, or ultra-broad with coverage over the whole range of interest.   This paper suggests the latter approach, and demonstrates that an air plasma driven by two-color laser pulses \cite{kim2007,karpowicz09,oh2013} produces suitable spectral content and pulse energy.  Suitability of the transverse mode is left for future work.

An alternative to the two-color air plasma scheme considered here would be to utilize advanced schemes of photonic down-conversion in crystals \cite{maidment2018,qu2019,wilson2019}.  An advantage of the air plasma is that it is simple and low cost, provided the titanium:sapphire drive pulse is already available.  Furthermore, scaling up the energy requires only that a thin sample of the ubiquitous beta barium borate (BBO) be increased in cross sectional area.

The specific plasma compression scenario that is the ultimate object of this work uses the Brookhaven National Laboratory BESTIA laser as the pump.  BESTIA uses CO$_2$ isotope mixtures to produce few picosecond pulse durations in various bands, with multi-joule pulse energy \cite{polyanskiy2017}.  As a nominal case, consider a 3 joule, 3 picosecond pump pulse with wavelength 10.6 $\mu$m.  Based on the parameter space study in \cite{johnson2017}, combined with the constraint of plasma length $\approx 1$ mm (due to gas jet dimensions), we choose focused spot size 500 $\mu$m, and plasma density $10^{18}$ cm$^{-3}$.  A three dimensional turboWAVE \cite{turbowave} simulation is carried out, in which the seed pulse from the two-color plasma is represented as a 100\% bandwidth, near single-cycle pulse, with 1 $\mu$J of energy.  In this scenario the amplified pulse is compressed to 0.13 picoseconds with 0.7 joules of energy.  A set of one dimensional runs, in which the seed pulse energy was varied between 0.01 $\mu$J and 100 $\mu$J, shows that the output parameters change by only tens of percent over four decades of seed pulse energy.  The particular turboWAVE module used is the cold relativistic fluid model, see Ref.~\cite{johnson2017}.

The remainder of this paper presents experiments and simulations of two-color laser excitation of air plasmas, wherein LWIR pulses are generated, with parameters suitable for seeding plasma compression of high energy CO$_2$ laser pulses.

\section{Experimental Setup}

\begin{figure}
\includegraphics[width=3in]{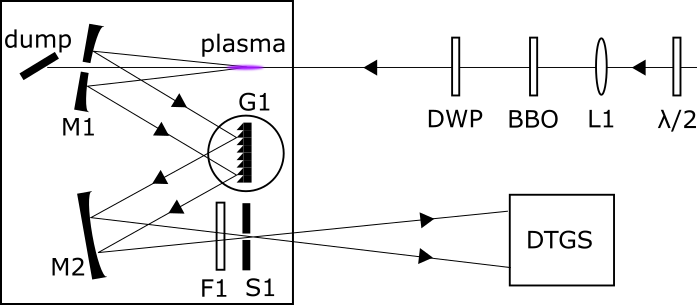}
\caption{Schematic of the experimental setup.  The near infrared drive pulse enters from the upper right.  Diverging rays are generated in the plasma.  Elements of the system are described in the text.}
\label{fig:setup}
\end{figure}

The setup used to generate and characterize the LWIR pulses is shown in Fig.~\ref{fig:setup}.  The experiments were carried out at the Brookhaven National Laboratory Accelerator Test Facility (ATF).  The driving near infrared laser produces up to 7.5 mJ at a center wavelength of 0.8 $\mu$m and pulse duration 55 fs.  The spot size radius measured at $1/e$ of the amplitude is about 3 mm, and the repetition rate is 240 Hz.  The driving laser pulse is polarized vertically, which is out of the page in Fig.~\ref{fig:setup}.  The driving pulse first encounters a half-waveplate, which can be used to refine or deliberately spoil the polarization.  The lens $L1$ focuses the pulse about 1 meter downstream.  Well upstream of the focus are the second harmonic generator (SHG) and the dual waveplate (DWP).  The SHG is a 100 micron thick type I BBO crystal, i.e., if the incident 0.8 micron pulse is vertically polarized, the 0.4 micron output is horizontally polarized.  The dual waveplate acts as a half-waveplate for 0.8 micron radiation, and a full waveplate for 0.4 micron radiation.  For optimal LWIR yield, it is rotated such that both colors are horizontally polarized upon output.  

The two-color pulse is focused into an enclosure containing a custom $4f$ grating spectrometer, where $f$ is the focal length of the curved mirrors M1 and M2.  The two-color drive pulse is focused at the entrance plane, i.e., $f=20$ cm upstream of M1.  The focused laser pulse is of sufficient intensity to photoionize air by multi-photon ionization, and simultaneously undergo significant Kerr self-focusing.  It is possible for a self-guided mode to develop, extending the region where the intensity is very high.  As a result, the mirror M1 can be damaged by the intense central core of the radiation.  This problem can be turned into an advantage by exploiting the fact that the LWIR radiation of interest is emitted into a cone angle that exceeds the divergence angle of the drive pulse \cite{oh2013}.  In particular, a 3 mm diameter hole in M1 serves to spatially filter out the drive pulse, while reflecting and focusing the LWIR radiation.  The drive pulse is dumped behind M1, while the LWIR radiation continues through the usual $4f$ spectrometer configuration, i.e., grating G1 positioned $f=20$ cm from M1, and mirror M2 positioned $f=20$ cm from G1.  Grating G1 has 75 grooves per mm.  Various filters $F1$ can be positioned at the output slit to help eliminate ambiguities arising from parasitic grating orders.  The deuterated triglycine sulfate (DTGS) detector enclosure is positioned 30 cm from the output slit, $S1$.  Inside the detector enclosure a parabolic collector focuses the radiation onto the DTGS sensor element.

The grating G1 is mounted on a rotation stage such that the axis of rotation lies on the grating surface.  The spectrometer is calibrated by matching four observed appearance angles of a 3.4 $\mu$m helium-neon laser to their theoretical values.  The DTGS sensor is calibrated using a commercial optical parameteric amplifier and difference frequency generation source.  Measurement of the DTGS signal as a function of grating angle then produces a calibrated multi-shot spectrum.  The optical elements in the spectrometer, along with the DTGS detector, have a weak dependence on wavelength through the LWIR range of interest.  On the other hand, the LWIR must propagate through just over 1 meter of ambient, humid air, which attenuates wavelengths outside the water window, i.e., outside the range of 8-14 microns.

\section{Second Harmonic Generation}

\begin{figure}
\includegraphics[width=3in]{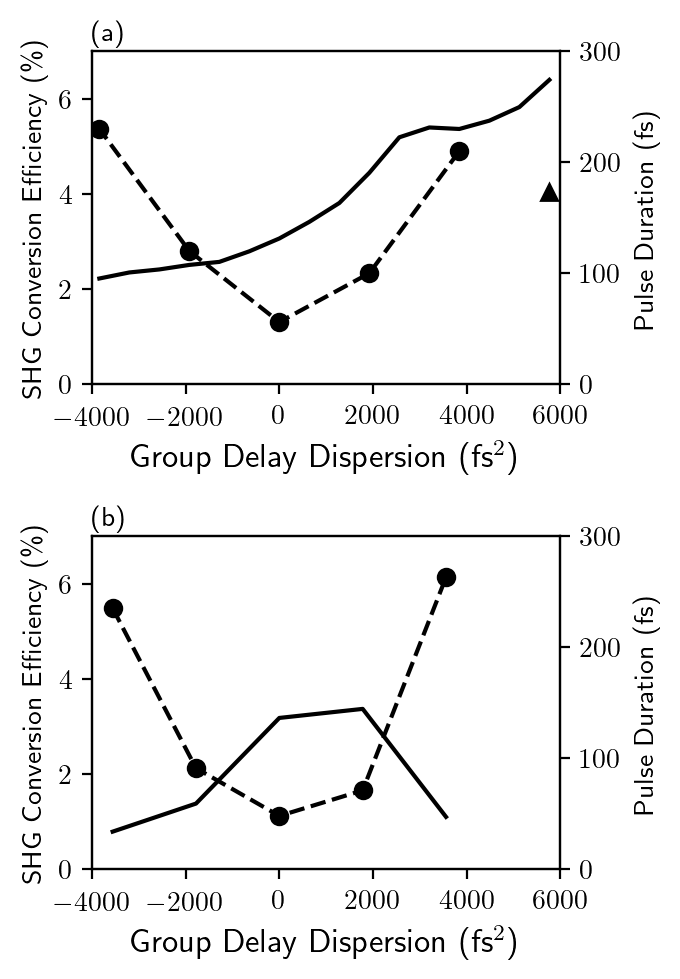}
\caption{SHG conversion efficiency (solid line) vs. GDD in the upstream pulse compressor based on (a) experimental measurement and (b) SNLO simulation with fixed phase mismatch.  The dots show the pulse duration for selected values of the GDD based on (a) measurement and (b) calculation.  The triangle shows a pulse duration measurement that is likely outside of the autocorrelator's useful range.}
\label{fig:shg}
\end{figure}

The LWIR generation mechansim herein depends crucially on SHG efficiency.  Figure~\ref{fig:shg}(a) shows the measured SHG conversion efficiency as a function of the group delay dispersion (GDD) associated with the pulse compressor.  The pulse duration is also shown, as defined by autocorrelation width divided by $2^{1/2}$.   Experimentally, it was verified that multi-photon absorption is not important by measuring $98\%$ transmission through the BBO crystal.  The intensity of the shortest pulse in the crystal is $\sim 700$ GW$/$cm$^2$.

Simulations using the Sandia Nonlinear Optics (SNLO) code \cite{SNLO}, in which a perfect lowest order Gaussian is perfectly phase matched, and higher order nonlinearities are neglected, predict conversion efficiency of $\approx 70\%$ for GDD=0, with symmetrical lowering of the efficiency for chirped pulses of either sign.  This is not consistent with the low efficiency and asymmetry in Fig.~\ref{fig:shg}(a).  One can recover a slight asymmetry about GDD=0 by accounting for uncompensated third and fourth order dispersion of the laser system, but the effect is too small to explain the observed SHG characteristics.

By imposing a fixed phase mismatch, the SNLO simulation can be made to resemble the experimental result.  This is illustrated in Fig.~\ref{fig:shg}(b), which shows the simulation analog of Fig.~\ref{fig:shg}(a).  The simulation includes an uncompensated third order dispersion of $-45000$ fs$^3$ and fourth order dispersion of $400000$ fs$^4$.  The radius of curvature of the pump wave phase fronts is -100 cm.  The fixed phase mismatch is $100$ rad$/$mm.  Such a mismatch, i.e., one that cannot be compensated by angle tuning, could be explained by partial coherence of the pump wave, or third order nonlinearities in the crystal.

The literature on SHG with $\sim 0.8$ $\mu$m, $\sim 50$ fs pulses, indicates that it is possible to achieve much higher conversion efficiency using more elaborate arrangements.  For example, Ref.~\cite{gobert2014} achieves $>65\%$ efficiency with $\sim 1$ mJ pulses by using noncollinear, chirped pumps with an optimized pulse-front tilt.  The intensity in the crystal is kept low, $<10$ GW$/$cm$^2$, by stretching the pulse to the picosecond range, and using a thicker crystal to make up for the reduced nonlinearity.  In order to use such a configuration for two-color laser-plasma interactions, one would have to separately re-compress the fundamental and second harmonic, and synchronously focus them, nearly collinearly, in the air.

\section{Experimental LWIR Spectra}

The spectra obtained during the experimental campaign are multi-shot spectra, obtained by acquiring data at various grating angles, with a 1 mm output slit.  In all cases the plotted detector signal represents the maximum signal detected over several shots.  The transverse dimensions of the source correspond to an effective input slit width in the tens of microns range.  The longitudinal extension of the source, based on simulation, is in the 2-3 centimeter range.  Given these parameters, the output slit width dominates the spectrometer resolution.  The dispersion of the spectrometer in the band of interest varies between .063 and .067 microns of wavelength per mm.  The nominal resolution with the 1 mm output slit is therefore $\approx 0.065$ microns.  This is well within the sampling interval of the data readout, corresponding to 0.4 microns of wavelength.

\begin{figure}
\includegraphics[width=3in]{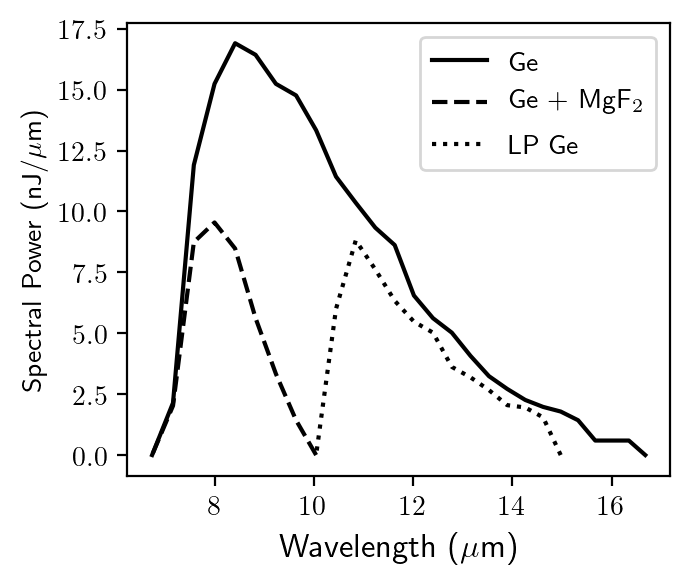}
\caption{Measured LWIR spectra through various filters, F1, in addition to 1 meter of ambient air}
\label{fig:spectrum}
\end{figure}

The spectra shown in Fig.~\ref{fig:spectrum} correspond to 6.3 mJ of drive pulse energy, with the shortest possible pulse duration, about 55 fs.  The three curves are obtained through different choices of the filter F1.  The purpose of varying the filter is to establish that parasitic grating orders do not pollute the spectrum, which is important in an environment where the radiation incident on the grating may have frequency components throughout a broad range.  The germanium filter is expected to have a fairly flat response throughout the interrogated wavelength range, and therefore represents the best estimate of the true spectrum.  The addition of MgF$_2$ gives the dashed curve, which suppresses wavelengths longer than about 8 microns, as expected.  The dotted line corresponds to a coated germanium wafer engineered to cut-on at 10.6 microns, as is indeed observed.  In all cases the spectrum is filtered by 1 meter of ambient air.

\begin{figure}
\includegraphics[width=3in]{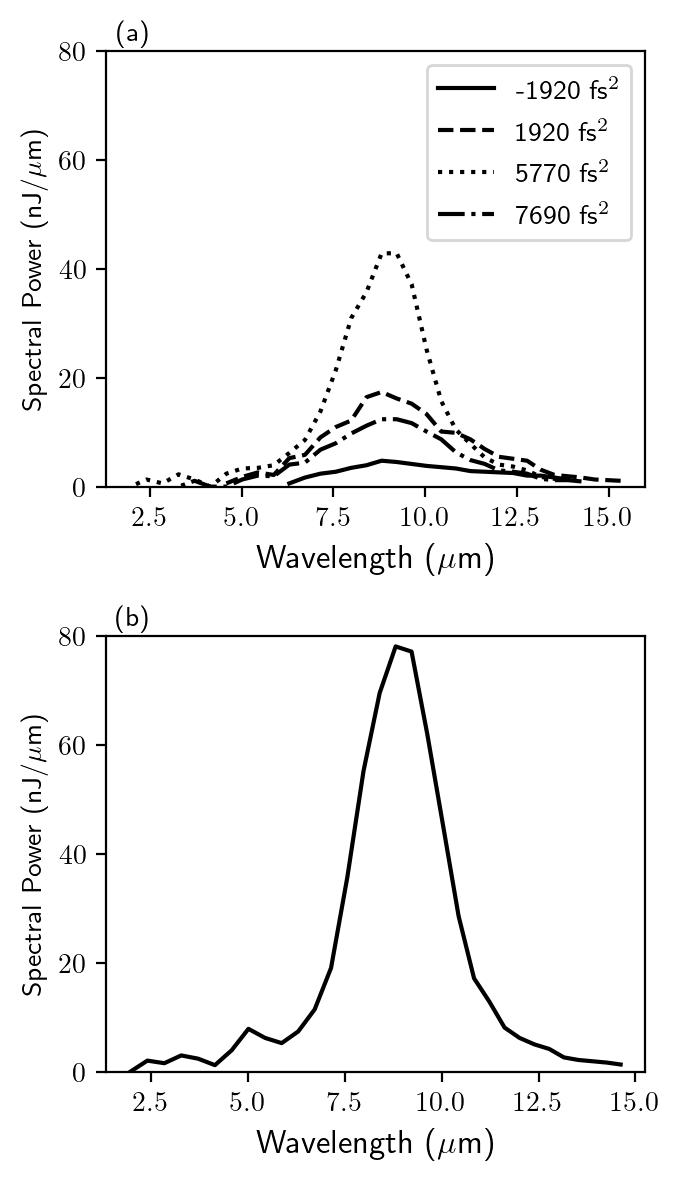}
\caption{Effect of drive pulse chirp and energy on LWIR spectra.  Panel (a) shows four spectra for four different values of the compressor GDD setting, holding drive pulse energy fixed at 6.3 mJ.  Panel (b) shows a spectrum with 7.5 mJ of drive pulse energy, and GDD of $5770$ fs$^2$.}
\label{fig:chirp}
\end{figure}

One expects the LWIR yield to depend on SHG efficiency.  We confirmed that the LWIR is extinguished by detuning the BBO phase matching angle by about 3 degrees.  The SHG efficiency is also affected by the compressor GDD, as seen in Fig.~\ref{fig:shg}.  The effect of GDD on the LWIR yield, holding pulse energy fixed, is shown in Fig.~\ref{fig:chirp}(a).  This dependency is complicated, because the chirp, pulse duration, and SHG efficiency all change with GDD.  Furthermore, the effective nonlinear refractive index of air involves a delayed rotational component, which tends to enhance self-focusing for longer pulses \cite{sprangle2002,zahedpour2015}.  Simulations presented below suggest that the latter effect can be dominant.

In order to test the limits of the given laser system, the total energy is tuned to the maximum possible value of 7.5 mJ, and the GDD is tuned to $5770$ fs$^2$ to optimize the SHG efficiency.  This produces the spectrum in Fig.~\ref{fig:chirp}(b).   The area under the curve is the pulse energy, 240 nJ.  By replacing the spectrometer grating with a mirror at the zero order angle, up to 335 nJ is directly measured with a pyrometer.  

\section{Simulations of LWIR Generation}

The dependence of the experimental LWIR yield on the compressor GDD requires detailed explanation.  For this purpose, simulations are carried out using the unidirectional pulse propagation equation \cite{couairon2011} (UPPE) module of SeaRay \cite{searay}.  The simulations are carried out in mixed coordinates, i.e., a single Cartesian polarization component is evolved on an axisymmetric cylindrical grid.  The dispersion relation for 40\% humid air is applied using the expression in \cite{voronin2017}.  This dispersion relation gives the real part of the susceptibility over a broad spectral range.  The nonlinear response of neutral air molecules is modeled as a constant nonlinear susceptibiltiy, $\chi^{(3)}$, which is varied by hand with pulse duration to approximate the effects of molecular rotations \cite{sprangle2002,zahedpour2015}.  Photoionization is evaluated in the tunneling limit, using the Coulomb corrected quasi-classical PPT theory \cite{perelemov66,perelemov67,perelemov67.2,perelemov68} , with an effective residual charge of $Z=0.53$, ionization potential $12.1$ eV, and gas density of $5.4\times 10^{18}$ cm$^{-3}$.  The former parameters are chosen to approximate ionization rates for molecular oxygen \cite{talebpour1999}.

The SeaRay code combines eikonal, paraxial, and UPPE models.  In the simulations presented here, a transform limited lowest order Gaussian pulse is created, with parameters chosen to match the experiment.  The eikonal model propagates the pulse through an ideal stretcher, harmonic generator, and lens.  The ideal stretcher applies a spectral phase shift $\beta(\omega-\omega_0)^2/2$, where $\beta$ is the GDD coefficient. The ideal harmonic generator of order $n$, efficiency $\eta$, and harmonic delay $\tau$, moves a fraction of spectral power, $\eta/n$, from a component $(\omega,{\bf k})$ with phase $\varphi$, to a new component $(n\omega,n{\bf k})$ with phase $n\varphi + n\tau\omega$.  In all the simulations herein $n=2$ and $\tau = 20$ fs.  The ideal lens has focal length 90 cm.  The UPPE portion of the model is initialized using the eikonal wave data evaluated 20 cm upstream of the geometric focus.  The phase difference between the second harmonic and fundamental is in general a complex function of the simulated optical elements.

\begin{table}[htp]
\caption{Simulated scan of GDD}
\begin{center}
\begin{tabular}{|c|c|c|c|}
\hline
Run & GDD, $\beta$ & SHG efficiency, $\eta$ & Nonlinear index, $n_2$  \\
 & (fs$^2$) & & (m$^2/$W) \\
\hline
\hline
1 & -4000 & .02 & $4.0\times 10^{-23}$ \\
2 & -2000 & .03 & $1.8\times 10^{-23}$ \\
3 & 0 & .04 & $1.0\times10^{-23}$ \\
4 & 2000 & .05 & $1.8\times 10^{-23}$ \\
5 & 4000 & .06 & $4.0\times 10^{-23}$ \\ 
\hline
\end{tabular}
\end{center}
\label{tab:params}
\end{table}%

\begin{figure}
\includegraphics[width=3in]{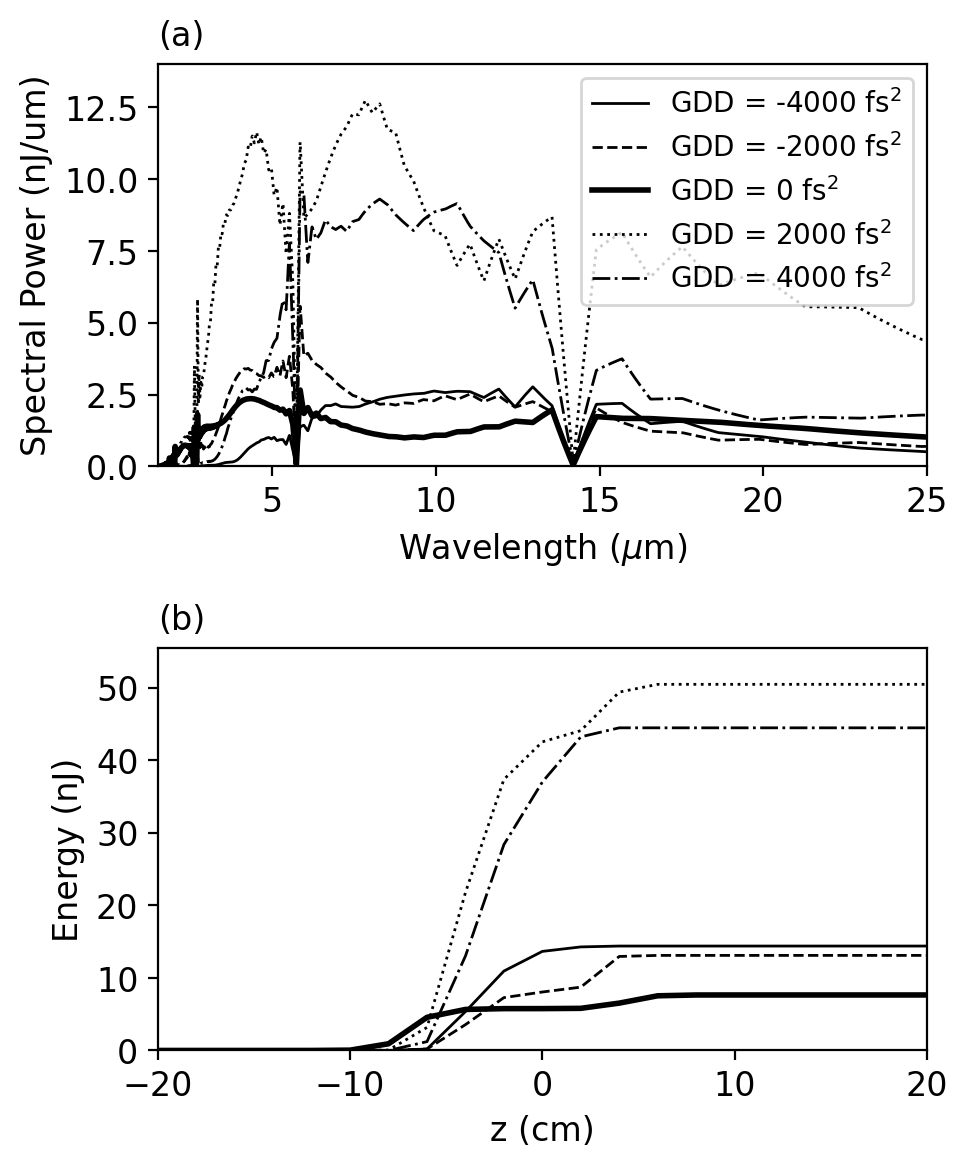}
\caption{Effect of drive pulse chirp on simulated LWIR spectra, showing (a) spectra, and (b) evolution of energy in the band $8$-$14$ $\mu$m.  The values of $\eta$ and $n_2$ are varied per Table~\ref{tab:params}.  The legend in (a) applies to (b).}
\label{fig:chirp-sim}
\end{figure}

Consider first a simulation scan over GDD where SHG efficiency and nonlinear refractive index are varied phenomenologically.  The SHG efficiency is taken as a simple linear function of GDD, while the nonlinear refractive index is taken to depend quadratically on GDD, with coefficients chosen to produce a range consistent with established measurements, such as those in Ref.~\cite{zahedpour2015}.  The particular values used are given in Table~\ref{tab:params}.  Pulse duration and power are computed in the usual way for a Gaussian pulse envelope.  The corresponding LWIR spectra are shown in Fig.~\ref{fig:chirp-sim}(a), and the evolution of energy in the band $8$-$14$ $\mu$m with propagation distance is shown in (b).  The most important characteristic of the experimental data is recovered, i.e., there is an optimum positive GDD which greatly enhances the LWIR yield (cf. Fig.~\ref{fig:chirp}).  The simulated and experimental LWIR yield are on the same order.  The simulated spectra have broader bandwidth than those observed in experiment.  This may be due in part to the fact that the spectrometer grating is optimized for 10 micron radiation.  Limitations of the model include simplified treatment of the molecular response, evaluation of photoionization in the tunneling limit, and assumption of a perfectly coherent Gaussian pump beam.

\begin{figure}
\includegraphics[width=3in]{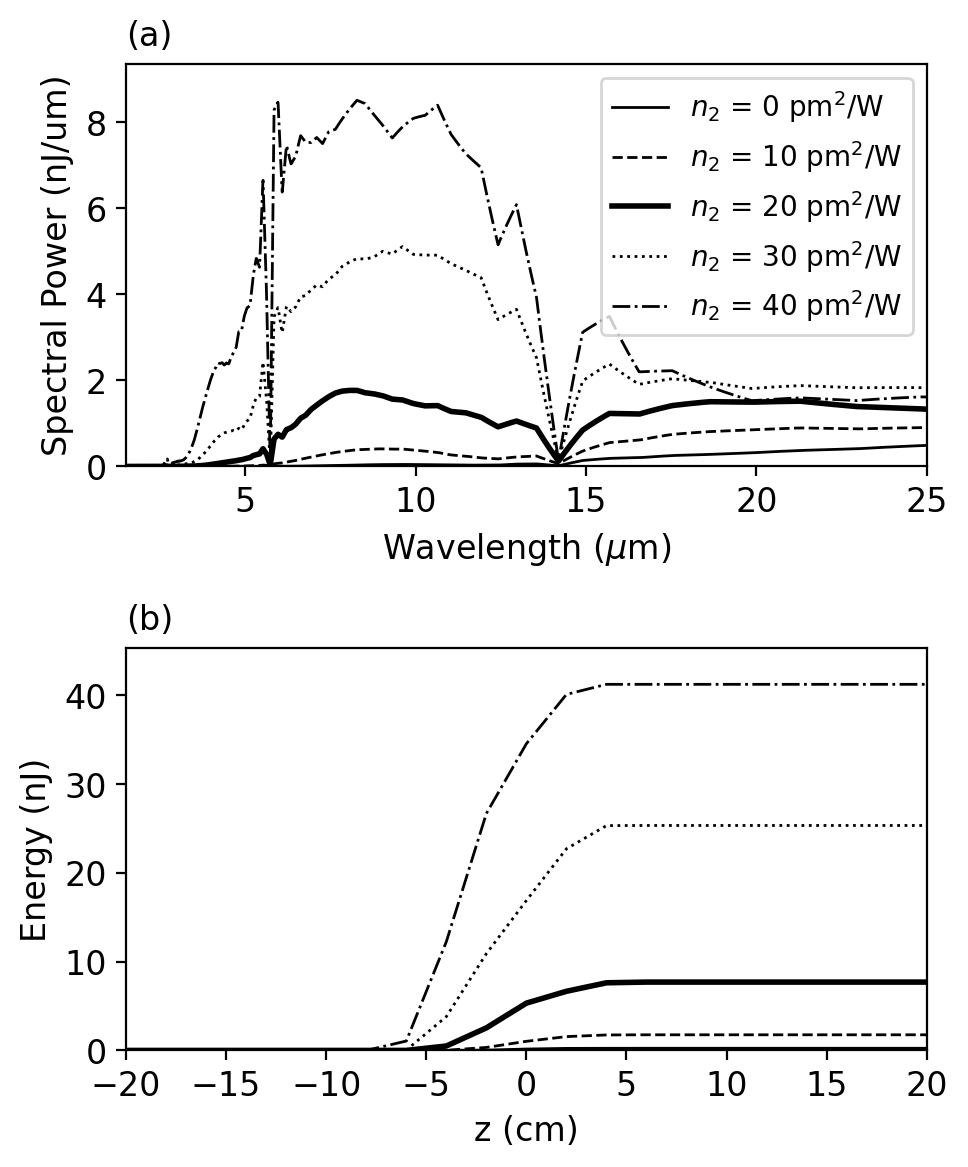}
\caption{Effect of Kerr nonlinearity on simulated LWIR spectra for $\eta = 5\%$ and GDD$=4000$ fs$^2$, showing (a) spectra, and (b) evolution of energy in the band $8$-$14$ $\mu$m.  The legend in (a) applies to (b).}
\label{fig:kerr-scan}
\end{figure}

The fact that a stretched pulse with a particular chirp enhances the LWIR is due in part to the enhanced SHG efficiency, but also due to the enhanced self focusing associated with an elevated $n_2$.  To illustrate the latter effect, Fig.~\ref{fig:kerr-scan} shows a set of simulations with GDD of $4000$ fs$^2$, SHG efficiency fixed at $\eta = 5\%$, and variable nonlinear refractive index, $n_2$.  The LWIR yield is a strong function of $n_2$.  This is because LWIR generation is due to ionization, which is a strong function of intensity, which is in turn elevated by Kerr induced self-focusing.

\begin{figure}
\includegraphics[width=3in]{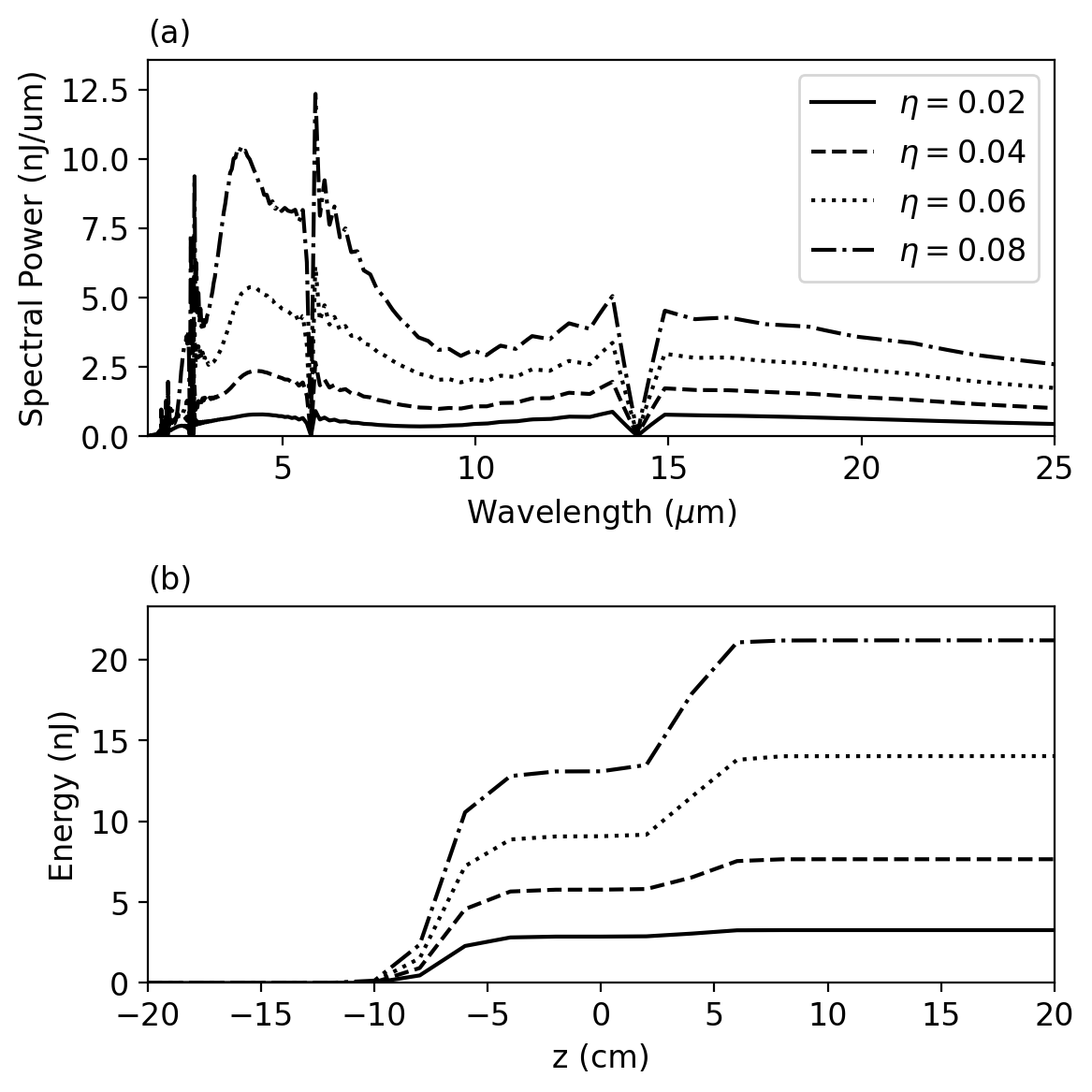}
\caption{Effect of second harmonic efficiency on simulated LWIR spectra for $n_2 =10^{-23}$ m$^2/$W and GDD$=0$, showing (a) spectra, and (b) evolution of energy in the band $8$-$14$ $\mu$m.  The legend in (a) applies to (b).}
\label{fig:eta-scan}
\end{figure}

Fig.~\ref{fig:eta-scan} shows the effect of varying SHG efficiency while holding $n_2$ fixed, with the shortest possible pulse (GDD=0).  The LWIR yield is approximately proportional to the SHG efficiency.  Furthermore, the spectral shape is insensitive to the SHG efficiency.  An interesting feature of the case with the shortest pulse is the secondary burst of LWIR generation around $z=5$ cm.  This secondary burst, due to Kerr re-focusing, is responsible for most of the radiation that appears in the band $3$-$5$ $\mu$m.

\begin{figure*}
\includegraphics[width=6.5in]{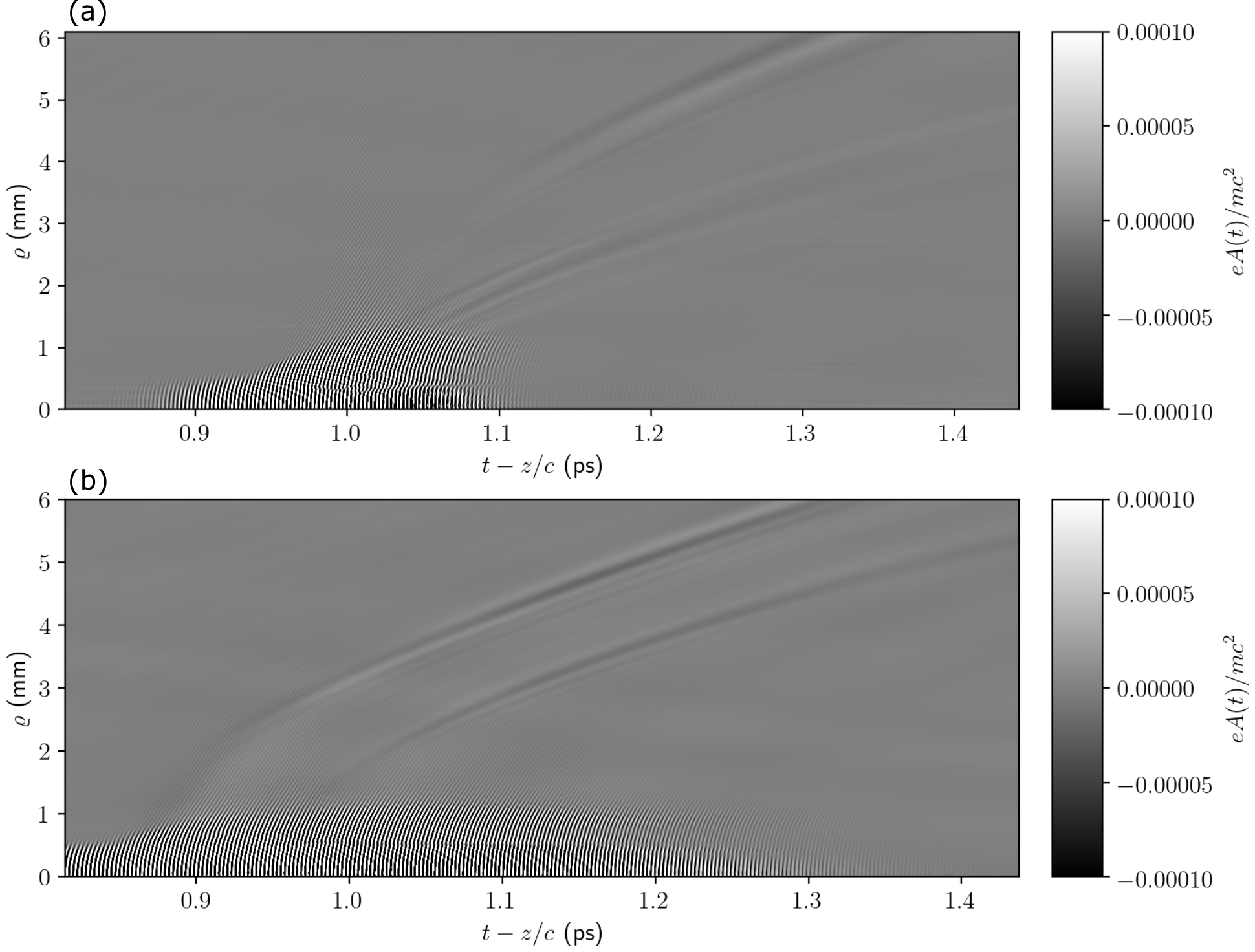}
\caption{Greyscale plot of $A(t,\varrho)$ at $z = 12$ cm, for (a) run 3, and (b) run 5 from Table~\ref{tab:params}.  The faint streaks appearing at $\varrho>1.5$ mm correspond to LWIR radiation.  The features nearer the axis correspond to the drive laser (zoom in to resolve carrier wave).}
\label{fig:time-domain}
\end{figure*}

Finally, Fig.~\ref{fig:time-domain} shows the total field in the time domain, $A(t,\varrho,z)$.  Here, $\varrho$ is the distance from the axis, and $z$ is held fixed at 12 cm downstream of the geometric focal point.  Panels (a) and (b) are for runs 3 and 5, respectively, from Table~\ref{tab:params}.  In both cases the drive pulse remains near the axis, while the LWIR separates radially due to larger divergence angle.  The LWIR emerges in two single-cycle bursts, seen as diagonal dark and light streaks.  In the short pulse case, the two bursts are generated at different z-positions, while in the long pulse case, both bursts are generated around the same z-position (but at different times).  This is why in the short pulse case the burst nearer the axis leads in time.

The consequence of double seed pulses on plasma compression depends on the separation between pulses.  Suppose the pump is strongly depleted, i.e., the compressor is efficient.  If the inverse of the pulse separation is well within the compressor bandwidth, the leading pulse will sweep up most of the pump energy, and the trailing pulse will produce only a small post-pulse.  If the inverse of the pulse separation greatly exceeds the compressor bandwidth, the compressor sees the two pulses as merged, but there is a large frequency mismatch.  In the case considered here, the inverse of the pulse separation is slightly greater than the compressor bandwidth, so that the two pulses are nearly merged, but there is a modest frequency mismatch.

\section{Conclusions}

Two-color laser pulses gently focused in air produce LWIR radiation suitable for seeding plasma compression in the LWIR regime.  The source requires only a few optical elements in addition to a several mJ class ultra-short pulse near-infrared drive laser.  The LWIR radiation covers the ``water window'' from $8$-$14$ $\mu$m.  This radiation may be useful for plasma compression with a CO$_2$ laser.  Remaining issues include the suitability of the transverse mode, and the pulse format in the time domain.

The LWIR spectra are most favorable when the drive pulse compressor is set to a substantial positive GDD.  This is due to the enhancement of the nonlinear refractive index for longer pulses, and the observed enhancement of the SHG efficiency for positive GDD.

\section{Acknowledgments}
This work was supported by the Office of Naval Research.  We thank M. Palmer for assisting with experimental arrangements.  We thank M. Polyanskiy and I. Pogorelsky for useful advice on diagnostic equipment during the experiments.  ATF operation is supported by the U.S. Department of Energy under contract DE-SC0012704.


\end{document}